\documentclass[superscriptaddress,secnumarabic,amssymb,amsmath,nobibnotes,aps,prd,showkeys,nofootinbib,preprint]{revtex4}

\usepackage{graphicx}
\usepackage{float}
\usepackage{bm}
\usepackage{amsmath}
\usepackage{amsfonts}
\usepackage{amssymb}
\usepackage{epstopdf}
\usepackage{natbib}%
\newcommand{\bee}{\begin{equation}}
\newcommand{\eee}{\end{equation}}
\newcommand{\eaa}{\end{eqnarray}}
\newcommand{\baa}{\begin{eqnarray}}
\def\ni{\noindent}

\usepackage{color}

\begin{document}

\title{Statistical approaches and the Bekenstein bound conjecture \\ in Schwarzschild black holes }

\author{Everton M. C. Abreu}\email{evertonabreu@ufrrj.br}
\affiliation{Departamento de F\'{i}sica, Universidade Federal Rural do Rio de Janeiro, 23890-971, Serop\'edica, RJ, Brazil}
\affiliation{Departamento de F\'{i}sica, Universidade Federal de Juiz de Fora, 36036-330, Juiz de Fora, MG, Brazil}
\affiliation{Programa de P\'os-Gradua\c{c}\~ao Interdisciplinar em F\'isica Aplicada, Instituto de F\'{i}sica, Universidade Federal do Rio de Janeiro, 21941-972, Rio de Janeiro, RJ, Brazil}
\author{Jorge Ananias Neto}\email{jorge@fisica.ufjf.br}
\affiliation{Departamento de F\'{i}sica, Universidade Federal de Juiz de Fora, 36036-330, Juiz de Fora, MG, Brazil}
\
%\pacs{04.60.Pp, 04.70.Dy, 05.20.-y}
\keywords{Bekenstein bound, Barrow entropy, Tsallis entropy, Kaniadakis entropy}
%%%%%%%%%%%%%%%%%%%%%%%%%%%%%%%%%%%%%%%%%%%%%%%%%%%%%%%%%%%%%%%%%%%%%%%%%%%%%%%%%%%%%%%%%%%%

%%%%%%%%%%%%%%%%%%%%%%%%%%%%%%%%%%%%%%%%%%%%%%%%%%%%%%%%%%%%%%%%%%%%%%%%%%%%%%%%%%%%%%%%%%
\begin{abstract}
\noindent One of the challenges of today's theoretical physics is to fully understand the connection between a geometrical object like area and a thermostatistical one like entropy, since we have theoretical proofs that the area behaves analogously like entropy does.  
The Bekenstein bound suggests a universal constraint for the entropy in a flat space region.
The Bekenstein-Hawking entropy of black holes satisfies the Bekenstein bound conjecture.
In this paper we have shown that when we use important non-Gaussian entropies, like the ones 
of Barrow, Tsallis and Kaniadakis to describe the Schwarzschild black hole, then the Bekenstein
bound conjecture seems to fail.
\end{abstract}
%%%%%%%%%%%%%%%%%%%%%%%%%%%%%%%%%%%%%%%%%%%%%%%%%%%%%%%%%%%%%%%%%%%%%%%%%%%%%%%%%%%%%%%%%%%%
\date{\today}

\maketitle
%%%%%%%%%%%%%%%%%%%%%%%%%%%%%%%%%%%%%%%%%%%%%%%%%%%%%%%%%%%%%%%%%%%%%%%%%%%%%%%%%%%%%%%%%%%%%%

\section{Introduction}

Bekenstein \cite{jdb} observed that one of the black hole features, i.e., its area, has traits that are similar to the characteristics of entropy.
Entropy, without a doubt, is one of the main quantities of thermodynamics.
As a matter of fact, the Hawking area theorem \cite{swh} states that in any classical scenario, the area $A$ does not decrease.
In other words, it acts exactly like the entropy does. In fact, it was found that the similarities between black holes physics and thermodynamics are rather broad.
Basically, there are laws of thermodynamics reinterpreted in terms of the thermal properties of black holes.
It is interesting to note that the entropy of a black hole can reveal its intrinsic quantum nature although it depends on a classical quantity that is the area 
of the event horizon.

A universal upper bound upon the entropy of a confined quantum system was also proposed by Bekenstein \cite{beke} and its mathematical formulation is given by

\bee
\label{bekenstein1}
S \leq \frac{2 \,\pi\, k_B\, R\, E}{\hbar \,c} \,,
\eee

\ni where $S$ denotes the entropy, $k_B$ is the Boltzmann constant, $E$ is the total energy and $R$ is the radius of a sphere that can comprise the given system.
Eq. (\ref{bekenstein1}) is known as the Bekenstein bound conjecture. We can note in Eq. (\ref{bekenstein1}) that for $\hbar \rightarrow 0$ we obtain the entropy S $\leq \infty $, 
meaning that the entropy of a localized system is unbounded from above in a classical regime. We can also observe that Eq. (\ref{bekenstein1}) does not contain the Newton constant $G_N$, pointing to the possibility of a Bekenstein bound conjecture not being restricted only to gravitational phenomena.   It is important to say that despite the fact that there are some counter-examples \cite{unruh,page,unruh2} showing some possible flaws, there are many examples \cite{beke2,beke3,beke4,beke5} that confirm Eq. (1) and the Bekenstein bound conjecture can be rigorously demonstrated by using conventional quantum mechanics 
and quantum field theory in flat spacetime \cite {cas}. Here we can mention that the generalized uncertainty principle is also used to derive the Bekenstein 
bound \cite{bg}. From now on we will adopt the natural unity system where
\bee
\label{uni}
k_B = \hbar = c = G = 1 \,.
\eee

\ni Using the values in Eq. (\ref{uni}) we can write Eq. (\ref{bekenstein1}) as
\bee
\label{bekenstein2}
S \leq 2 \,\pi\, R\, E \,.
\eee

\ni  To derive a specific form of Eq. (\ref{bekenstein2}), compatible with a black hole system, we begin with the Schwarzschild metric which is given by
\bee
\label{sch}
ds^2 = \left( 1 - \frac{2 M}{R} \right) dt^2 - \left( 1 - \frac{2 M}{R} \right)^{-1} dr^2 - r^2 d \Omega^2 \,.
\eee

\ni The black hole solution leads to 
\bee
\label{bhs}
R = 2 M \,.
\eee 

\ni We will assume that the radius $R$ of a sphere which encloses a given system and appears in Eq. (\ref{bekenstein2}) is the Schwarzschild radius. 
Using Eq. (\ref{bhs}) and considering that $ E = M $, thus Eq. (\ref{bekenstein2}) can be rewritten as
\bee
\label{bekenstein3}
S \leq \pi R^2 \,.
\eee

\ni % We will denote Eq. (\ref{bekenstein3}) as the Bekenstein bound conjecture (BBC).%
We can see that relation in Eq. (\ref{bekenstein3}) is saturated when $ S = \pi R^2 $ for the Schwarzschild black holes whose entropy is given by $ S = A_H/4 $ where $ A_H $ is the horizon area given by $ 4 \pi R^2 $.

The aim of this paper is to show that the Bekenstein bound conjecture is not satisfied when Barrow, Tsallis, and Kaniadakis entropies are used to describe black hole 
thermostatistics.
So, in order to clarify the structure of our paper this work is organized as follows: in Sections 2, 3 and 4, the entropies of Barrow, Tsallis and Kaniadakis respectively, are
used to investigate the black hole thermodynamics and to verify if the Bekenstein bound conjecture is valid for each model.
%we will use the Barrow entropy to describe the black holes thermodynamics 
%and verify if the Bekenstein bound conjecture is valid. In Section III Tsallis statistics is used to examine the Bekenstein bound. In Section IV we will use the Kaniadakis 
%approach to investigate the Bekenstein bound conjecture.
Section V is dedicated to the conclusions and final words. 

\section{Barrow entropy}

In a seminal paper \cite{barrow-1}, Barrow investigated the conditions and what quantum gravitational effects it could cause on the tangled structure of the black hole horizon, 
inspired by the Covid-19 virus schematic depiction. This complex structure presents a finite volume but it has an infinite (or finite) area \cite{saridakis-0} and therefore it  
can modify the actual horizon area of a black hole. This leads to a new entropy area relation which is
\bee
\label{barrow} 
S_B\,=\,\bigg(\frac{A}{A_o} \bigg)^{ 1+ \frac{\Delta}{2} }, 
\eee

\ni where $A$ is the usual horizon area  and $A_o$, the Planck area. Although it is a sort of Tsallis nonextensive entropy \cite{tsallis 1,varios-4,tcir}, it is important to note that Barrow's entropy differs from the usual quantum corrected ones \cite{saridakis-1,saridakis-2,svar,nos,nos-2} and the ones with logarithmic corrections \cite{kc}. 

From Eq. (\ref{barrow}) we can see that this quantum gravitational perturbation is represented by the new parameter $\Delta$.  There are some representative  values for $\Delta$.  
For instance, when $\Delta= 0$, the horizon is in its most basic form, i.e., the Bekenstein-Hawking entropy.
When $\Delta=1$ we have the so called maximal deformation which means that Barrow's entropy depicts an extended fractal black hole.   This implies that this entropy can be much larger than the standard Schwarzschild value.   Hence, we can expect that the standard features for black holes, i.e., no hair entropy bound, etc., will not obey the underlying forms.

Considering the horizon area as $A = 4 \pi R^2$ and $A_o = 4 $, then Eq. (\ref{barrow}) can be read as

\bee
\label{barrowu}
S_B\,=\, \left(\pi R^2 \right)^{1+\frac{\Delta}{2}} \,.
\eee

\ni Assuming that Barrow's entropy describes the black holes thermodynamics, thus using Eq. (\ref{barrowu}) we can write the Bekenstein conjecture, 
Eq. (\ref{bekenstein3}), as

\bee
\label{bekenb}
S_B \, \leq \, S_B^\frac{2}{2+\Delta} \,\,,
\eee

\ni which is the Bekenstein bound conjecture written in the context of Barrow entropy. However, we can see that Eq. (\ref{bekenb}) 
is not satisfied for $ S_B > 1$. We have defined the ratio $R_B$ as $ R_B \equiv S_B^{2/(2+\Delta)}/S_B $. This definition
can be simplified in the form $R_B = 1/{S_B^\frac{\Delta}{2+\Delta}}$.
In Fig. 1 we have plotted the ratio $ R_B $ as a function of $\Delta$ for $ S_B = 2$ where we can observe that $R_B < 1$ as long as
$S_B > 1$ within the valid interval of $\Delta$ which is $0 \leq \Delta \leq 1$ \cite{barrow-1}.

\begin{figure}[H]
	\centering
	\includegraphics[height=5.cm,width=7.cm]{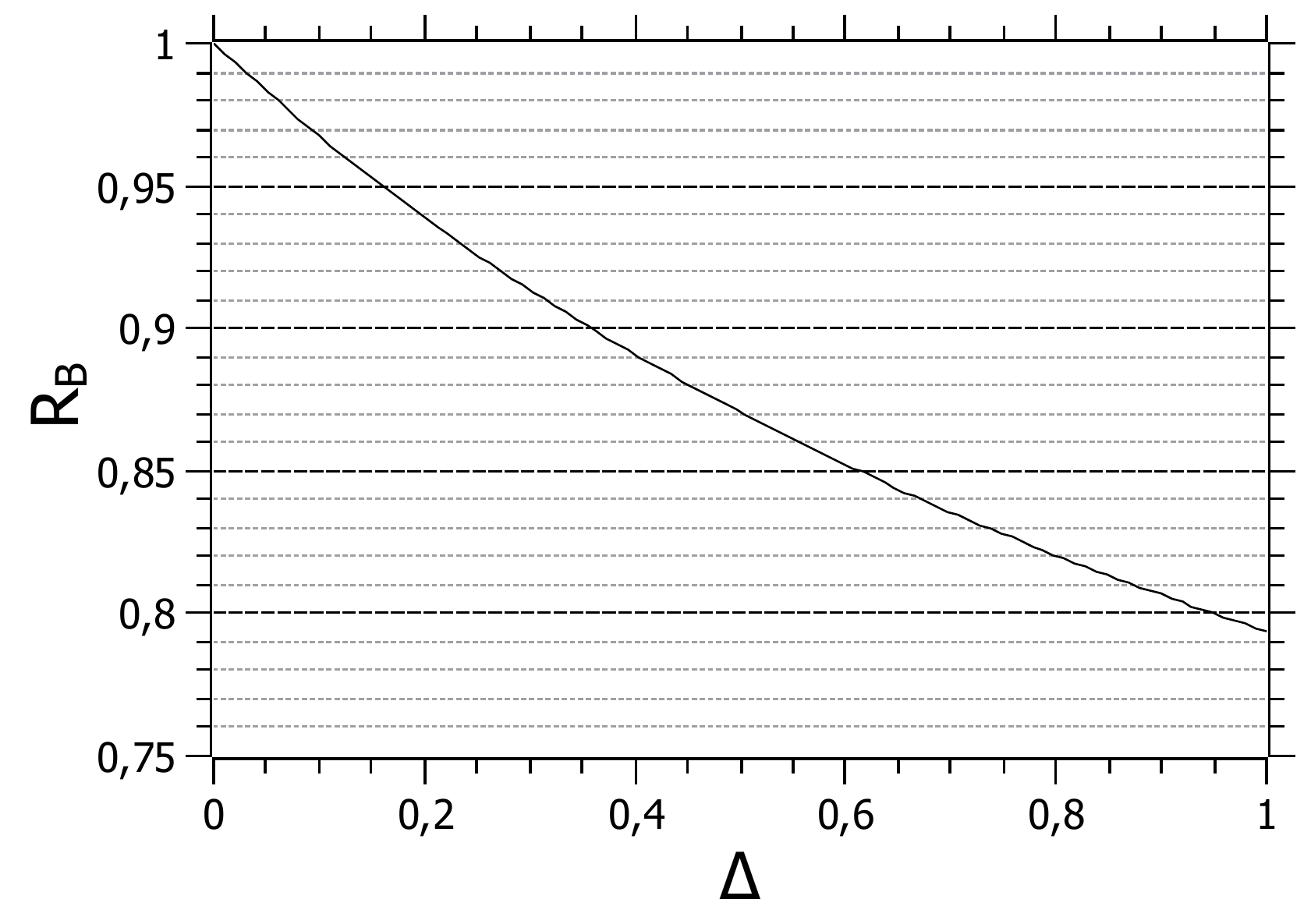}
	\caption{Values of  $R_B = 1/{S_B^\frac{\Delta}{2+\Delta}}$ as a function of $\Delta$ for $ S_B = 2$.}
	\label{barrowb}
\end{figure}

\ni From Fig. 1 we can see two different results: the first one is that for $\Delta = 0$ we have $R_B=1$. 
This result says that we have an equality in Eq. (\ref{bekenb}). The second result is that when $\Delta$ increases, the value of $R_B$, 
which is always less than one, decreases even more. This result shows that the Bekenstein bound conjecture is not satisfied when Barrow's 
entropy is used to describe the black holes thermodynamics for $ S_B > 1$.

%Here it is worth mentioning that %
As we mentioned before we can observe that Barrow's entropy describes a fractal structure of a black hole 
horizon \cite{barrow-1}. As a consequence we have that the Barrow entropy can be significantly higher than the usual Bekenstein-Hawking entropy. Therefore, from 
the considerations presented above, it is possible to consider that the square of the horizon radius has the form $R^{2+\Delta}$ and then
the horizon radius has the form $R^{1+\frac{\Delta}{2}}$. Consequently, a modified Bekenstein conjecture could have the following form

\bee
\label{fractal}
S \leq 2\pi R^{1+\frac{\Delta}{2}} E \,.
\eee

\ni Making $\Delta = 0$ in Eq. (\ref{fractal}) we recover the usual Bekenstein bound conjecture, Eq. (\ref{bekenstein2}). Considering that $E=M$ and $M=2R$ then we have

\bee
\label{fractal2}
S \leq \pi R^{2+\frac{\Delta}{2}} \,.
\eee

\ni Assuming that Barrow's entropy describes the black hole thermodynamics, thus using Eq. (\ref{barrowu}) we can write the modified Bekenstein conjecture, 
Eq. (\ref{fractal}), as

\begin{eqnarray}
\label{bekenbf}
S_B \leq \pi^{- \frac{\Delta}{4}} S_B^{\frac{4+\Delta}{2(2+\Delta)}} \,,
\end{eqnarray}

\ni which is the modified Bekenstein bound conjecture written in the context of Barrow entropy.
In Fig. 2 we have plotted the ratio $ R_{BM} \equiv S_B^{2/(2+\Delta)}/S_B = 
\pi^{- \frac{\Delta}{4}} S_B^{-\frac{\Delta}{2(2+\Delta)}}  $ as a function of $\Delta$ for $ S_B = 2$.

\begin{figure}[H]
	\centering
	\includegraphics[height=5.cm,width=7.cm]{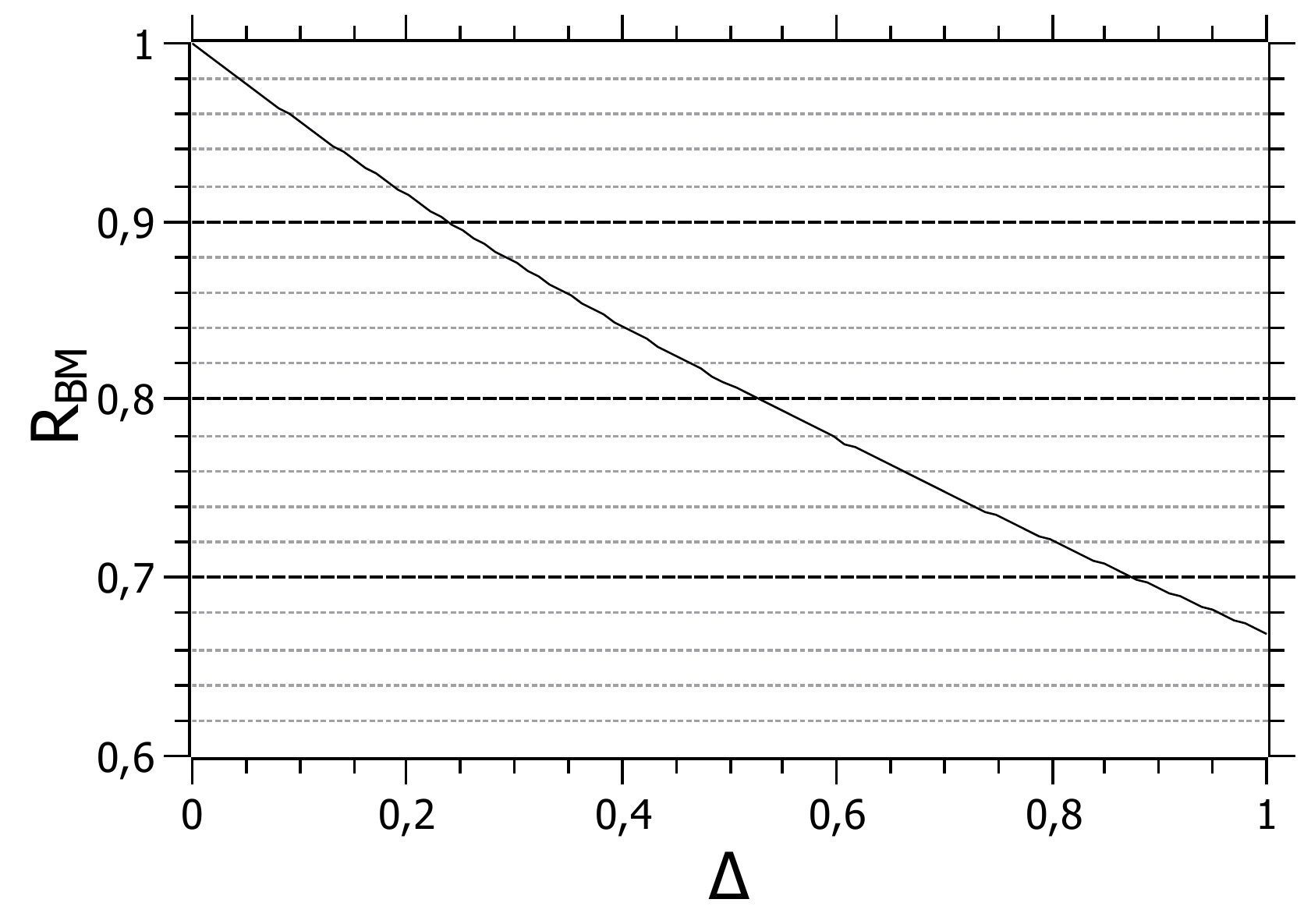}
	\caption{Values of $  R_{BM} = 
		\pi^{- \frac{\Delta}{4}} S_B^{-\frac{\Delta}{2(2+\Delta)}}$ as a function of $\Delta$ for $ S_{BM} = 2$.}
	\label{barrowb}
\end{figure}

\ni Just as we have observed in Fig. 1, from Fig. 2 we can see two distinguished results: The first one is that it is easy to see that $R_{BM} = 1$ for $\Delta = 0$.
This result states that we have the equality in Eq. (\ref{bekenbf}). The second result is that when $\Delta$ increases, the value of $R_{BM}$, 
which is always less than one, decreases even further. This result shows that a modified Bekenstein bound conjecture, Eq. (\ref{bekenbf}), 
is not satisfied when Barrow's entropy is used again to describe the black holes thermodynamics for $S_{BM}=2$.

\section{Tsallis statistics}

A nonextensive (NE) entropy is defined by Tsallis' statistics \cite{tsallis 1}, which is an extension of Boltzmann-Gibbs (BG) statistical theory.
The Tsallis entropy formula is given by

\begin{eqnarray}
	\label{tsa}
	S_T = \frac{1-\sum_{i=1}^W p_i^q}{q-1} \;\;\;\;\; \left(\sum_{i=1}^W p_i=1\right) \,,
\end{eqnarray}

\ni where $p_i$ is the probability of a system to exist within a microstate, $W$ is the total number of configuration (microstates) and $q$, known as the Tsallis parameter or NE parameter in the current literature, is a real parameter that measures the degree of nonextensivity. The definition of entropy in Tsallis statistics  brings the standard properties of positivity, equiprobability, concavity and irreversibility. This proposal has been successfully applied to a variety of physical systems. For example,
we can mention the Levy-type anomalous diffusion \cite{levy}, turbulence in a pure electron-electron plasma \cite{celia} and gravitational systems \cite{many1,mora,nos1t,nos2t,nonextensive,ko}.
It is worth noting that Tsallis thermostatistics formalism considers BG statistics to be a special case in the limit $q\rightarrow 1$  where the usual additivity of entropy can be retrieved. In the microcanonical ensemble, where all the states have the same probability, Tsallis' entropy reduces to \cite{tsallis 1}

\begin{eqnarray}
	\label{tsam}
	S_T = \frac{W^{1-q}-1}{1-q} \,,
\end{eqnarray}

\ni where, in the limit $ q \rightarrow 1$, we recover the usual BG entropy formula, $ S = \ln\, W $.

 Considering that the BG entropy describes, at first, the Bekenstein-Hawking entropy, thus we have the following relation

\begin{eqnarray}
	\label{eqw}
\ln W = \pi R^2 \,. 
\end{eqnarray}

\ni From Eq. (\ref{eqw}) we have

\begin{eqnarray}
\label{eqw2}
  W = e^{\pi R^2} \,. 
\end{eqnarray}

\ni Now, using  Eq. (\ref{eqw2}) into Tsallis entropy, Eq. (\ref{tsam}), we find

\begin{eqnarray}
	\label{tsabh}
	S_T = \frac{e^{\lambda \pi R^2}-1}{\lambda} \,,
\end{eqnarray}

\ni where $\lambda \equiv 1-q$. Here it is worth noting that when we use Tsallis 
entropy to describe the thermodynamics of black holes the result is that 
the heat capacity can be positive for certain values of the mass of the black hole
unlike the negative value obtained when Boltzmann-Gibbs describes the same system.
This important outcome indicates that Tsallis entropy can lead black holes to be 
thermodynamically stable. For more details see reference \cite{me}.

Then, supposing that the Tsallis entropy describes the black holes thermostatistics and using Eq. (\ref{tsabh}) 
we can write the Bekenstein bound, Eq. (\ref{bekenstein3}), as

\begin{eqnarray}
  \label{tsabbc}
 S_T \, \leq \frac{1}{\lambda} \, \ln \left( 1 + \lambda \, S_T \right)\,.
\end{eqnarray}

\ni  Eq.(\ref{tsabbc}) is the Bekenstein bound conjecture formulated in the framework of Tsallis entropy. In order to verify 
if the inequality, Eq. (\ref{tsabbc}), is satisfied we have plotted in Fig. 3 the ratio defined as

\begin{eqnarray}
\label{rtsallis}
R_T \equiv \frac{\ln \left( 1 + \lambda \, S_T \right)}{\lambda S_T} \,, 
\end{eqnarray} 

\ni for usual values of $\lambda$ that is $\lambda \geq 0$ and $ S_T = 2$.

\begin{figure}[H]
	\centering
	\includegraphics[height=7.cm,width=8.cm]{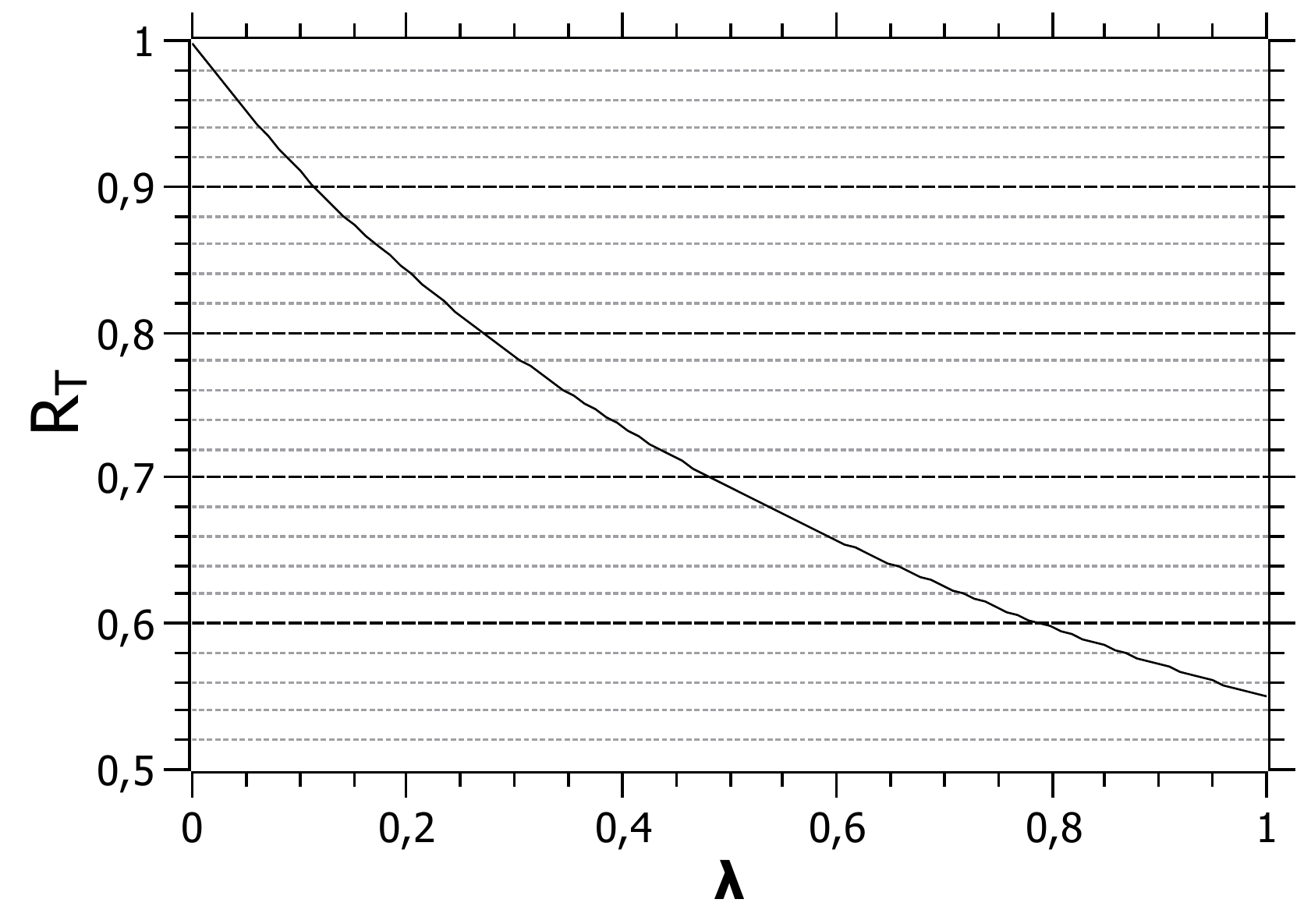}
	\caption{Values of $ R_T \equiv   \ln \left( 1 + \lambda \, S_T\right)/(\lambda S_T) $ as a function of $\lambda$ for $ S_T = 2$.}
	\label{tsallisb}
\end{figure}

\ni From Fig. 3 we can note two distinguished results: the first is that for $\lambda = 0$ we have $R_T=1$.
This property can be verified analytically using the power series expansion of logarithm for $\lambda \approx 0$
in the numerator of 
Eq. (\ref{rtsallis})  which gives the following approximation: $\ln \left( 1 + \lambda \, S_T\right) \approx \lambda \, S_T$.
According to this result we have that $R_T =1$ and therefore we have the equality in Eq. (\ref{tsabbc}). 
The second is that the value of $ R_T $, which is always less than one, decreases as 
the $\lambda$ parameter grows. Therefore, the Bekenstein bound conjecture is not satisfied when the thermodynamics of 
black holes is described by Tsallis entropy.

\section{Kaniadakis statistics}

The well-known Kaniadakis statistics \cite{kani}, also known as a $kappa-$ statistics, generalizes the usual BG statistics by first formulating  the $kappa$- entropy as

\begin{eqnarray}
	\label{kani}
	S_\kappa = -\sum_i^W \frac{p_i^{1+\kappa} -  p_i^{1-\kappa}}{2\kappa}  \,,
\end{eqnarray}

\ni which recovers the BG entropy at the limit $\kappa \rightarrow 0$. It is important to mention here that the $\kappa$-entropy satisfies the properties relative to concavity, additivity and extensivity. When used in many specific contexts, the $kappa$-statistics has been quite successful. As an example we can cite cosmic rays \cite{kani2} and gravitational and cosmological systems \cite{nos2,nos2k,tkh}. Using the microcanonical ensemble definition, where all the states have the same probability, Kaniadakis' entropy reduces to \cite{kani}

\begin{eqnarray}
	\label{kanim}
	S_\kappa = \frac{W^\kappa-W^{-\kappa}}{2\kappa} \,,
\end{eqnarray}

\ni where, in the limit $\kappa \rightarrow 0$, we recover the usual BG entropy formula, $S = \ln \, W$.

Considering again that the BG entropy describes, at first, the black hole thermodynamics, then we have the same relations used in Tsallis statistics section above which are Eqs. (\ref{eqw}) and (\ref{eqw2}).
Substituting Eq. (\ref{eqw2}) into Kaniadakis entropy, Eq. (\ref{kanim}), we have

\begin{eqnarray}
	\label{kanibbc1}
	S_\kappa = \frac{e^{\kappa \pi R^2}-e^{-\kappa \pi R^2}}{2\kappa} \,.
\end{eqnarray}

\ni  Assuming that the Kaniadakis entropy describes the black holes thermostatistics, thus using Eq. (\ref{kanibbc1}) we can write the Bekenstein bound conjecture, Eq. (\ref{bekenstein3}), as

\begin{eqnarray}
	\label{kanibbc2}
	S_\kappa \, \leq \frac{1}{\kappa} \, \ln \left( \kappa \, S_\kappa  + \sqrt{ 1 + \kappa^2 \, S_\kappa^2}     \right)        \,.
\end{eqnarray}

\ni Eq.(\ref{kanibbc2}) is the Bekenstein bound conjecture formulated in the context of Kaniadakis entropy. To verify if the inequality in Eq. (\ref{kanibbc2}) is satisfied we have plotted in Fig. 4 the ratio $R_\kappa$ defined as

\begin{eqnarray}
	\label{rkaniadakis}
R_\kappa \equiv \frac{\ln \left( \kappa \, S_\kappa + \sqrt{ 1 + \kappa^2  \, S_\kappa^2} \right)}{\kappa S_\kappa}	 \,, 
\end{eqnarray} 

%$  R_\kappa \equiv \left( \frac{1}{\kappa} \, \ln \left( \kappa \, S_\kappa + \sqrt{ 1 + \kappa^2  \, S_\kappa^2} \right) \right)  /S_\kappa $%
\ni for usual values of $\kappa$ that is $\kappa \geq 0$ and $ S_\kappa = 2$.

\begin{figure}[H]
	\centering
	\includegraphics[height=5.cm,width=7.cm]{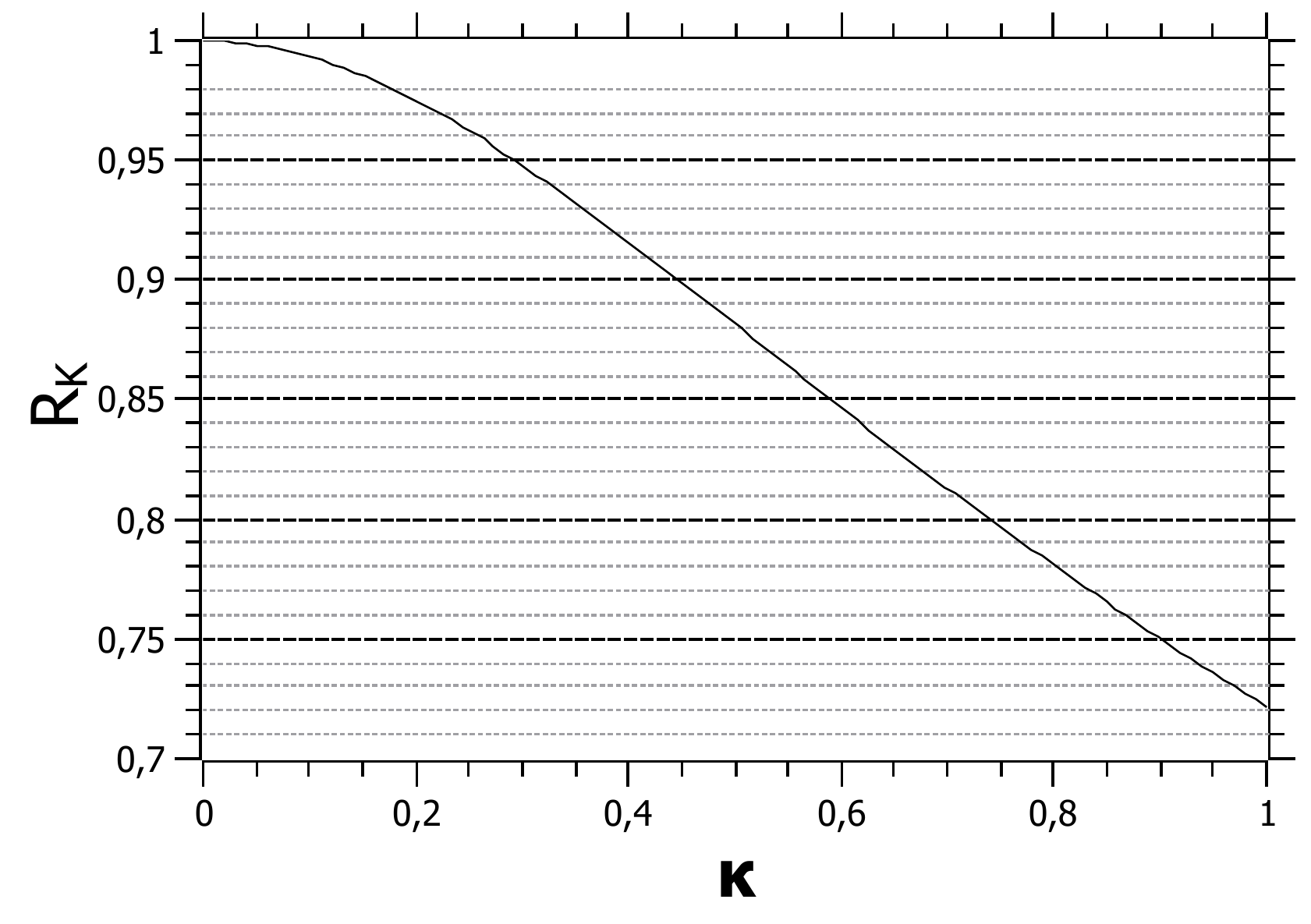}
	\caption{Values of $  R_\kappa \equiv  \ln \left( \kappa \, S_\kappa + \sqrt{ 1 + \kappa^2  \, S_\kappa^2} \right)  /(\kappa S_\kappa)  $ 
	as a function of $\kappa$ for $ S_\kappa = 2$.}
	\label{kaniadakisb}
\end{figure}

\ni From Fig. 4 we can observe two significant results: the first one is that for $\kappa = 0$ we have $R_\kappa=1$.
This feature can be verified analytically using the power series expansions of logarithm and square root for
$\kappa \approx 0$ in the numerator of Eq. (\ref{rkaniadakis}) which yields the following approximation: $ \ln \left( \kappa S_\kappa + 
\sqrt{ 1 + \kappa^2  \, S_\kappa^2} \right) \approx \kappa \, S_\kappa$.                                      
Consequently we have that $R_\kappa = 1$ and therefore we have the equality in Eq. (\ref{kanibbc2}). The second result is that 
when the $\kappa$ parameter increases, the value of $R_\kappa$, which is always less than one, decreases.
This result shows that the Bekenstein bound conjecture is not satisfied when the Kaniadakis entropy is used to describe the black holes thermodynamics.

\section{Conclusions}

It is well known that some extensions of the Boltzmann-Gibbs entropy, which we call non-gaussian entropies,
have been quite successful when applied to different models.
Therefore, it becomes relevant to know whether the Benkenstein bound conjecture remains valid when 
black holes are described by 
non-gaussian entropies. 
The presence of Planck's constant $\hbar$ in its formulation indicates that the Bekenstein bound conjecture can be an important inequality in 
quantum systems.
However, when the entropies of Barrow, Tsallis and Kaniadakis are used to describe black holes systems, this quantum inequality is not followed. 
This important result is based on the behavior of Eqs. (\ref{bekenb}), (\ref{tsabbc}) and (\ref{kanibbc2}).
As a perspective for future work, 
we plan to investigate possible Bekenstein bound violations in cosmological models.

\section*{Acknowledgments}

\ni We thank the anonymous Referee for useful comments that help us to upgrade this work.  Jorge Ananias Neto thanks CNPq (Conselho Nacional de Desenvolvimento Cient\'ifico e 
Tecnol\'ogico), Brazilian scientific support federal agency, for partial financial support, CNPq-PQ, Grant number 307153/2020-7.

\end{document}